\documentclass[a4paper,11pt,reqno]{article}
\usepackage{a4wide}

\usepackage{amsmath,amsfonts,amssymb}
\usepackage[T1]{fontenc}
\usepackage[p,osf]{cochineal}
\usepackage[varqu,varl,var0]{inconsolata}
\usepackage[scale=.95,type1]{cabin}
\usepackage[vvarbb]{newtxmath}
\usepackage[cal=boondoxo]{mathalfa}
\usepackage[mathscr]{euscript}
\usepackage{bbold} 

\usepackage{pict2e}

\makeatletter
\DeclareRobustCommand{\loplus}{\mathbin{\mathpalette\dog@lsemi{+}}}
\DeclareRobustCommand{\lotimes}{\mathbin{\mathpalette\dog@lsemi{\times}}}
\DeclareRobustCommand{\roplus}{\mathbin{\mathpalette\dog@rsemi{+}}}
\DeclareRobustCommand{\rotimes}{\mathbin{\mathpalette\dog@rsemi{\times}}}

\newcommand{\dog@rsemi}[2]{\dog@semi{#1}{#2}{-90,90}}
\newcommand{\dog@lsemi}[2]{\dog@semi{#1}{#2}{270,90}}
\newcommand{\dog@semi}[3]{%
  \begingroup
  \sbox\z@{$\m@th#1#2$}%
  \setlength{\unitlength}{\dimexpr\ht\z@+\dp\z@\relax}%
  \makebox[\wd\z@]{\raisebox{-\dp\z@}{%
    \begin{picture}(1,1)
    \linethickness{\variable@rule{#1}}
    \roundcap
    \put(0.5,0.5){\makebox(0,0){\raisebox{\dp\z@}{$\m@th#1#2$}}}
    \put(0.5,0.5){\arc[#3]{0.5}}
    \end{picture}%
  }}%
  \endgroup
}
\newcommand{\variable@rule}[1]{%
  \fontdimen8  
  \ifx#1\displaystyle\textfont3\else
    \ifx#1\textstyle\textfont3\else
      \ifx#1\scriptstyle\scriptfont3\else
        \scriptscriptfont3\relax
  \fi\fi\fi
}
\makeatother

\usepackage{lettrine}

\usepackage{nicefrac}
\usepackage{setspace}
\usepackage{datetime}
\usepackage[nosort]{cite}

\usepackage{upgreek}

\usepackage{comment}

\usepackage[euler]{textgreek}
\usepackage[breaklinks=true]{hyperref}

\hypersetup{
			colorlinks=true,
			urlcolor=blue,
			citecolor=magenta,
			linkcolor=blue,
     }

\let\oldsqrt\sqrt
\def\sqrt{\mathpalette\DHLhksqrt}
\def\DHLhksqrt#1#2{%
\setbox0=\hbox{$#1\oldsqrt{#2\,}$}\dimen0=\ht0
\advance\dimen0-0.2\ht0
\setbox2=\hbox{\vrule height\ht0 depth -\dimen0}%
{\box0\lower0.4pt\box2}}

\newcommand{\RNum}[1]{\uppercase\expandafter{\romannumeral #1\relax}}

\author{
  \begin{minipage}{.97\linewidth}
    \vspace{1cm}
       \begin{center}
      \begin{small}
      \textbf{David Rivera-Betancour}$^1$\footnote{david.rivera-betancour@polytechnique.edu}\,\, and 
      \textbf{Matthieu Vilatte}$^{1,2}$\footnote{matthieu.vilatte@polytechnique.edu}
              \end{small}
    \end{center}
    \vspace{0.5cm}
    \hspace{2.4cm}\begin{minipage}{.7\linewidth}
\begin{center}     {\it \begin{footnotesize}
\hbox{\kern-1.8cm\vbox{\vskip0cm
 \begin{itemize}
  \item[$^1$]Centre de Physique Th\'eorique -- CPHT\\ 
        Ecole Polytechnique, CNRS\footnote{\emph{Centre National de la Recherche Scientifique}, Unit\'e Mixte de Recherche UMR 7644.}\\
        Institut Polytechnique de Paris\\
        91128 Palaiseau Cedex, France                        
                           \vskip0.25cm
      \end{itemize}}
\kern-3.2cm\vbox{
\begin{itemize}
  \item[$^2$]Division of Theoretical Physics\\School of Physics\\
  Aristotle University of Thessaloniki\\ 
  54124 Thessaloniki, Greece
      \end{itemize}
      \vskip0.cm
}}
     \end{footnotesize}}
\end{center}
    \end{minipage}
    \vspace{0.5cm}\begin{minipage}{.7\linewidth}
     \end{minipage}
  \end{minipage}
}

\title{\vspace{2.5cm}
 \boldmath \begin{LARGE}
    \textbf{\textsc{Revisiting the Carrollian Scalar Field}}
  \end{LARGE} \unboldmath
}

\date{}

\begin{document}


\begin{titlepage}
\maketitle
\thispagestyle{empty}

 \vspace{-12.cm}
  \begin{flushright}
  CPHT-RR022.042022\\
  \end{flushright}
 \vspace{12.cm}

\begin{center}
\textsc{Abstract}\\  
\vspace{1. cm}	
\begin{minipage}{1.0\linewidth}

We investigate the (conformally coupled) scalar field on a general Carrollian spacetime in arbitrary dimension. The analysis discloses electric and magnetic dynamics. For both, we provide the energy and the momenta of the field, accompanied by their conservation equations. We discuss the conservation and non-conservation properties resulting from the existence of conformal isometries and the associated charges. We illustrate those results for a scalar field propagating on the null boundary of four-dimensional Ricci-flat Robinson--Trautman spacetimes.

\end{minipage}
\end{center}


\end{titlepage}

\onehalfspace

\begingroup
\hypersetup{linkcolor=black}

\endgroup

\lettrine[lines=2, lhang=0.33, loversize=0.25]{T}{he Carroll group} was discovered in the seminal works of L\'evy-Leblond  \cite{Levy} and Sen Gupta  \cite{SenGupta}. Its first application, due to Henneaux, appeared fourteen years later  \cite{Henneaux:1979vn} and it took almost fifty years for Carrollian physics to emerge as a full-blown research area, ranging from differential geometry to holographic duality. Carrollian physics is meant to embrace phenomena occurring on a Carrollian spacetime, such as hydrodynamics or, at a more fundamental level, field dynamics. The simplest field is a scalar and it has received some attention \cite{CM1, Bagchi:2019xfx, Bagchi:2019clu, Gupta:2020dtl, henneaux2021carroll, dutch, aritra2022, Hao}. 

The aim of the present note is to present the dynamics of a (conformally coupled) scalar on a general Carrollian manifold, tame and illustrate scattered results, and unify two distinct and complementary approaches. The first relies on Carrollian structures and diffeomorphism invariance.  The second consists in reaching Carrollian geometry and dynamics from a pseudo-Riemannian relative at vanishing speed of light. The set of features we address includes: (\romannumeral1) electric vs. magnetic dynamics; (\romannumeral2) action and equations of motion; (\romannumeral3) energy, momentum and their conservation; (\romannumeral4) isometries  and N\oe ther's theorem. The basic technical tools are  listed in the appendix.

\lettrine[lines=2, lhang=0.33, loversize=0.25]{C}{arroll structures} were introduced in \cite{Duval:2014uoa, Duval:2014uva, Duval:2014lpa} (see also \cite{Bekaert:2014bwa,Bekaert:2015xua, Hartong:2015xda, Morand:2018tke, Ciambelli:2019lap, Herfray:2021qmp}). They consist of a $d+1$-dimensional manifold $\mathscr{M}= \mathbb{R} \times \mathscr{S}$  equipped with a degenerate metric and a vector field, the kernel of the metric. 
For concreteness, we will adopt coordinates $( t, \mathbf{x})$ and degenerate metrics of the form
\begin{equation}   
\label{cardegmet}
\text{d}\ell^2=a_{ij}( t, \mathbf{x}) \text{d}x^i \text{d}x^j,\quad i,j\ldots \in \{1,\ldots,d\}
\end{equation}
with kernel generated by 
\begin{equation}   
\label{kert}
\upupsilon
= \frac{1}{\Omega}\partial_t,
\end{equation}
which defines a \emph{field of observers}.
This coordinate system is adapted to the space/time splitting, which is in turn respected by Carrollian diffeomorphisms 
 \begin{equation}
\label{cardifs} 
t'=t'(t,\mathbf{x})\quad \text{and} \quad \mathbf{x}^{\prime}=\mathbf{x}^{\prime}(\mathbf{x}).
\end{equation}

The Carrollian manifold incorporates an \emph{Ehresmann connection}, which is the background gauge field  
$\pmb{b}=b_i \text{d}x^i$ appearing in the dual form of the field of observers \eqref{kert}, defined such as $\upmu(\upupsilon)=-1$:
\begin{equation}   
\label{kertdual}
\upmu=-\Omega \text{d}t +b_i \text{d}x^i,
\end{equation}
the \emph{clock form} ($\Omega$ and $b_i$ depend on $t$ and $\mathbf{x}$). The vector fields dual to the forms $\text{d}x^i$ are 
\begin{equation}
\label{dhat}
\hat\partial_i=\partial_i+\frac{b_i}{\Omega}\partial_t.
\end{equation}
As shown in the appendix, they transform covariantly under  Carrollian diffeomorphisms \eqref{cardifs}. A Carroll structure (strong definition) is also equipped with a torsionless and metric-compatible connection. This is not unique, due to the degeneracy of the metric. We use here the connection inherited from the parent relativistic spacetime. 

A Carroll structure endowed with metric \eqref{cardegmet} and clock form \eqref{kertdual} is naturally reached in the Carrollian limit ($c\to 0$) of a pseudo-Riemannian spacetime  $\mathscr{M}$ in Papapetrou--Randers gauge 
\begin{equation}
\label{carrp}
\text{d}s^2 =- c^2\left(\Omega \text{d}t-b_i \text{d}x^i
\right)^2+a_{ij} \text{d}x^i \text{d}x^j,
\end{equation}
where all functions are $x$-dependent with $x\equiv(x^0=ct,\mathbf{x})$. The connection we use on the Carrollian side is given in the appendix, Eqs.  \eqref{dgammaCar} and \eqref{dgammaCartime} . These are parts of the Levi--Civita connection attached to \eqref{carrp}, and decomposed in powers of $c$.

\lettrine[lines=2, lhang=0.33, loversize=0.25]{T}{he dynamics of scalar fields} on an arbitrary Carrollian spacetime, limited to two-derivative kinetic terms encompasses two distinct situations dictated by Carrollian covariance. Their Lagrangian densities read:
\begin{eqnarray}
\label{Le}
\mathcal{L}_{\text{e}}
&=&\frac{1}{2 }\left(\frac{1}{\Omega}\partial_t \Phi\right)^2 - V_\text{e}(\Phi),\\
\label{Lm}
\mathcal{L}_{\text{m}}
&=&-\frac{1}{2} a^{ij}\hat\partial_i \Phi\hat\partial_j \Phi  - V_\text{m}(\Phi),
\end{eqnarray}
and enter the Carrollian action $ S_{\text{C}} = \int_\mathscr{M}  \text{d}t \,\text{d}^{d}x \sqrt{a} \Omega \mathcal{L}$. The indices ``e'' and ``m'' stand for \emph{electric} and \emph{magnetic}.  They refer to the origin of these actions in the parent relativistic theory \cite{henneaux2021carroll, dutch}. Indeed starting from a relativistic scalar field on a Papapetrou--Randers background \eqref{carrp}
\begin{equation}
\label{relS}
S=- \int_\mathscr{M}  \text{d}t \,\text{d}^{d}x \sqrt{-g}  \left(\frac{1}{2} g^{\mu\nu} \partial_\mu \Phi \partial_\nu \Phi  + V(\Phi)\right),
\end{equation}
and assuming 
\begin{equation}
\label{expV}
V(\Phi)= \frac{1}{c^2} V_\text{e}(\Phi)+  V_\text{m}(\Phi)+ \, \mathcal{O}\left(c^2\right),
\end{equation}
we find
\begin{equation}
\label{relSexp}
S= \frac{1}{c^2}S_{\text{e}} + S_{\text{m}}+ \, \mathcal{O}\left(c^2\right)
\end{equation}
with $S_{\text{e}}$ and $S_{\text{m}}$ the Carrollian actions with Lagrangian densities \eqref{Le} and \eqref{Lm}. 
The existence of an expansion \eqref{expV} for the original relativistic potential in powers of $c^2$ is a bona fide assumption, necessary to reach two actions invariant under Carrollian diffeomorphisms \eqref{cardifs}.\footnote{The actions  associated with the $\mathcal{O}\left(c^2\right)$ terms are non-dynamical as no kinetic term appears at this order. This will be illustrated in the subsequent analysis of a conformally coupled scalar, see Eq. \eqref{confpotexp}.}

Due to the form of the metric  \eqref{carrp}, and to its subsequent behaviour under Carrollian diffeomorphisms, the decomposition of any relativistic tensor as a (usually truncated) Laurent expansion, provides a Carrollian tensor for each term.\footnote{Phrased in more mathematical terms, the expansion in powers of $c^2$, amounts to reducing the representations of the full diffeomorphism group, with respect to the Carrollian diffeomorphism subgroup.} If we insist in reaching a single Carrollian tensor at vanishing $c$, then an appropriate rescaling by some power of $c^2$ is necessary -- in order  e.g. to select one out of  two options, if only two options are available as in the above scalar-field action (see \cite{henneaux2021carroll}, were this procedure is illustrated in Hamiltonian formalism and for flat spacetime). 

An insightful scalar potential for a relativistic curved spacetime in $d+1$ dimensions is the following:
\begin{equation}
\label{confpot}
V(\Phi)= \frac{d-1}{8d}R\Phi^2.
\end{equation}
For a scalar field $\Phi$ of weight  $w=\frac{d-1}{2}$, this is a conformal coupling. Indeed, the relativistic energy--momentum tensor for \eqref{relS} with \eqref{confpot} has the form ($\nabla_{\mu} \Phi =\partial_{\mu} \Phi $)
\begin{eqnarray}
\label{varrelT}
&T_{\mu\nu}=-\frac{2}{\sqrt{-g}}\frac{\delta S}{\delta g^{\mu\nu}}=
   \nabla_{\mu}  \Phi \nabla_{\nu} \Phi - \frac{1}{2} g_{\mu \nu}  \nabla_{\alpha}  \Phi  \nabla^{\alpha}  \Phi + \frac{d - 1}{4d} \left( G_{\mu \nu}\Phi^{2} + g_{\mu \nu}  \Box \Phi^{2} - \nabla_{\mu}  \nabla_{\nu}  \Phi^{2}\right)
\nonumber \\
&=\mathscr{D}_{\mu}  \Phi\mathscr{D}_{\nu} \Phi - \frac{1}{2} g_{\mu \nu} \mathscr{D}_{\alpha}  \Phi \mathscr{D}^{\alpha}  \Phi + \frac{d - 1}{4d} \left( \left( \mathscr{R}_{(\mu \nu)}-\frac{ \mathscr{R}}{2}g_{\mu \nu}  \right)\Phi^{2} + g_{\mu \nu}  \mathscr{D_\alpha}\mathscr{D}^\alpha \Phi^{2} - \mathscr{D}_{(\mu} \mathscr{D}_{\nu)}  \Phi^{2}\right),
\end{eqnarray}
where\footnote{We thank Konstantinos Siampos for a useful discussion on this topic.} $G_{\mu \nu}$ is the Einstein tensor,  $\mathscr{R}_{\mu \nu}$ and $ \mathscr{R}$ the  Weyl-covariant Ricci  and scalar defined in the appendix (Eqs. \eqref{curlRic} and \eqref{curlRc}), together with the Weyl-covariant derivative $\mathscr{D}_\mu$. This energy--momentum tensor is traceless when $\Phi$ is on-shell, and Weyl-covariant of weight $d-1$. The action is Weyl-invariant (up to boundary terms\footnote{Equation \eqref{varreleoms} is also simply $-\Box\Phi+ \frac{d-1}{4d} 
R\Phi=0$.}), whereas the equations of motion can be recast readily with Weyl-covariant attributes:
\begin{equation}
\label{varreleoms}
-\mathscr{D}_{\mu}\mathscr{D}^{\mu}\Phi+ \frac{d-1}{4d} 
\mathscr{R}\Phi=0.
\end{equation}
As a consequence of diffeomorphism invariance, the energy--momentum tensor obeys a Weyl-covariant 
conservation equation, when the field $\Phi$ is on-shell:
\begin{equation}
 \label{conconTJ} 
\nabla_\mu T^{\mu\nu}=\mathscr{D}_\mu T^{\mu\nu}=0.
\end{equation}

The interest for studying relativistic conformally coupled scalar fields is originally found in inflationary models of cosmology.\footnote{See e.g. \cite{cosmo} where more references are displayed.} On the Carrollian side the motivation is entrenched in the attempts to generalize the gauge/gravity holographic correspondence for asymptotically flat spacetimes, where the boundary is null infinity, i.e. a Carrollian manifold par excellence. 

Inserting inside \eqref{confpot}  the Carrollian decomposition of $R$ as displayed in the appendix Eq. \eqref{reducR},  leads to
\begin{equation}
\label{confpotexp}
V(\Phi)= \frac{1}{c^2} V_\text{e}(\Phi)+  V_\text{m}(\Phi) + c^2  V_\text{nd}(\Phi)  
\end{equation}
with
\begin{eqnarray}
V_\text{e}(\Phi)&=& \frac{d-1}{8d}\left(\frac{2}{\Omega}\partial_t \theta+\frac{1+d}{d}\theta^2+\xi_{ij}\xi^{ij}\right)\Phi^2
,\\
V_\text{m}(\Phi)&=& \frac{d-1}{8d}\left( \hat{r}-2\hat{\nabla}_{i}\varphi^{i}-2\varphi^{i}\varphi_{i}\right)\Phi^2,
\\
V_\text{nd}(\Phi)
&=& \frac{d-1}{8d}\varpi_{ij}\varpi^{ij}\Phi^2.
\end{eqnarray}
In the last expression the index ``nd'' stands for ``non-dynamical.'' The reason is that when the expression \eqref{confpotexp} of the potential is used in the relativistic action \eqref{relS}, it produces the Carrollian electric and magnetic  actions -- with some boundary terms dropped here\footnote{On the relativistic side we find: $\frac{1}{2} g^{\mu\nu} \partial_\mu \Phi \partial_\nu \Phi  + \frac{d-1}{8d}R\Phi^2=\frac{1}{2} \mathscr{D}^\mu \Phi  \mathscr{D}_\mu \Phi  + \frac{d-1}{8d} \mathscr{R}\Phi^2- \frac{d-1}{4\sqrt{-g}} 
\partial_\mu \left(\sqrt{-g} A^\mu\Phi^2
\right)
$.} 
\begin{eqnarray}
S_\text{e}&=& \int  \text{d}t \,\text{d}^{d}x \sqrt{a} \Omega \left(\frac{1}{2}\left(\frac{1}{\Omega}\hat{\mathscr{D}}_{t}\Phi\right)^2- \frac{d-1}{8d} 
\xi_{ij}\xi^{ij}\Phi^2\right)
\label{elaction}
,\\
S_\text{m}&=& \int  \text{d}t \,\text{d}^{d}x \sqrt{a} \Omega \left( -\frac12\hat{\mathscr{D}}_{i}\Phi\hat{\mathscr{D}}^{i}\Phi- \frac{d-1}{8d} 
\hat{\mathscr{R}}\Phi^2\right)
,
\label{magaction}
\end{eqnarray}
as well as a third one  $ S_{\text{nd}} =- \int  \text{d}t \,\text{d}^{d}x \sqrt{a} \Omega  \frac{d-1}{8d}\varpi_{ij}\varpi^{ij}\Phi^2$, which has no kinetic term for $\Phi$.\footnote{These results coincide with those obtained for $d=2$ 
in Ref. \cite{Gupta:2020dtl}, where the authors proceed with a thorough investigation of the possible Weyl-compatible terms. The kinetic terms of the electric and magnetic actions, \eqref{elaction} and \eqref{magaction}, can also be compared to the corresponding results of \cite{henneaux2021carroll}. They also agree up to the magnetic constraint introduced in Ref. \cite{henneaux2021carroll}, which would read here $\Pi^{i}_\text{m}=0$ (see \eqref{magenmom}). The latter guarantees the invariance of the action under local Carrollian boosts, which we have not required a priori -- Carrollian invariance features here the covariance under  Carrollian diffeomorphisms \eqref{cardifs} of a theory defined on a Carrollian spacetime \eqref{cardegmet} and \eqref{kert}.

} The Carrollian equations of motion for the two non-trivial cases are as follows:
\begin{eqnarray}
\label{elec-eom}
\frac{1}{\Omega}\hat{\mathscr{D}}_{t}\frac{1}{\Omega}\hat{\mathscr{D}}_{t}\Phi+ \frac{d-1}{4d} 
\xi_{ij}\xi^{ij}\Phi&=&0\quad \text{electric,}\\
\label{magn-eom}
 -\hat{\mathscr{D}}_{i}\hat{\mathscr{D}}^{i}\Phi+ \frac{d-1}{4d} 
\hat{\mathscr{R}}\Phi&=&0\quad \text{magnetic,}
\end{eqnarray}
where the detailed expressions for the derivatives and Carrollian tensors are available in the appendix. These equations are Weyl-covariant of weight $w=\frac{d+1}{2}$.

\lettrine[lines=2, lhang=0.33, loversize=0.25]{E}{nergy and momenta}  are part of the agenda when discussing field dynamics. These are  conjugate variables to the geometric data, as is $T_{\mu\nu}$ in \eqref{varrelT} for a relativistic theory, and inherit their conservation from the Carrollian diffeomorphism invariance. In Carrollian geometries there is no energy--momentum tensor, but instead an \emph{energy--stress tensor} $\Pi^{ij}$, an \emph{energy flux} $\Pi^{i}$ and an \emph{energy density} $\Pi$, defined as \cite{CM1,Chandrasekaran:2021hxc, BigFluid}:
\begin{eqnarray}
\label{carvarstren} 
\Pi^{ij}&=&\frac{2}{\sqrt{a} \Omega}\frac{\delta S_{\text{C}} }{\delta a_{ij}},\\
\label{carvarencur}
\Pi^{i}&=&\frac{1}{\sqrt{a} \Omega}\frac{\delta S_{\text{C}} }{\delta b_i},\\
\label{carvarenden}
\Pi&=&-\frac{1}{\sqrt{a}}\left(\frac{\delta S_{\text{C}} }{\delta \Omega}+\frac{b_i}{\Omega}\frac{\delta S_{\text{C}} }{\delta b_i}\right)
 \end{eqnarray}
with conformal weights $d+3$, $d+2$ and $d+1$. Requiring Weyl invariance for the action translates into
\begin{equation}
\label{car-conf-cond}
\Pi_i^{\hphantom{i}i}=\Pi,
\end{equation} 
valid on-shell (as the tracelessness of the relativistic energy--momentum tensor).

A \emph{momentum} $P_i $ (weight $d$) is also defined but is not conjugate to a geometric variable. It enters the conservation equations that mirror the Carrollian diffeomorphism invariance. For Weyl-invariant dynamics these are \cite{BigFluid}:
\begin{eqnarray}
\frac{1}{\Omega}\hat{\mathscr{D}}_t\Pi
+\hat{\mathscr{D}}_i \Pi^{i}
+\Pi^{ij}\xi_{ij}&=&0,
 \label{carEcon} 
\\
\hat{\mathscr{D}}_i \Pi^{i}_{\hphantom{i}j}+2\Pi^{i}\varpi_{ij}+ \left(\frac{1}{\Omega}\hat{\mathscr{D}}_t \delta^i_j +\xi^{i}_{\hphantom{i}j}\right) P_i &=&0.
  \label{carGcon}
\end{eqnarray}
Conservation equations are satisfied when the field $\Phi$ is on-shell, and this allows to determine the momentum. 

Using Eqs. \eqref{carvarstren}, \eqref{carvarencur} and  \eqref{carvarenden}, we obtain  the following energy and momenta for the Carrollian electric and magnetic actions: 
\begin{eqnarray}
\label{elenmom}
&&\begin{cases}
\Pi^{ij}_\text{e}=\frac{a^{ij}}{2}\left(\frac{1}{\Omega}\hat{\mathscr{D}}_t\Phi\right)^2+\frac{d-1}{4d}\left(\frac{1}{\Omega}\hat{\mathscr{D}}_t\left(\xi^{ij}\Phi^2\right)-a^{ij}\left(\frac12\xi_{lk}\xi^{lk}\Phi^2+\frac{1}{\Omega}\hat{\mathscr{D}}_t\frac{1}{\Omega}\hat{\mathscr{D}}_t\Phi^2\right)\right)
\\
\Pi^{i}_\text{e}=0
\\
\Pi_\text{e}=\frac{1}{2}\left(\frac{1}{\Omega}\hat{\mathscr{D}}_t\Phi\right)^2-\frac{d-1}{8d}\xi_{ij}\xi^{ij}\Phi^2,
\end{cases}
\\
\label{magenmom}
&&
\begin{cases}
\Pi^{ij}_\text{m}=\hat{\mathscr{D}}^i\Phi\hat{\mathscr{D}}^j\Phi-\frac{a^{ij}}{2}\hat{\mathscr{D}}_l\Phi\hat{\mathscr{D}}^l\Phi+\frac{d-1}{4d}\left(\left(\hat{\mathscr{R}}^{(ij)}-\frac{\hat{\mathscr{R}}}{2}a^{ij}\right)\Phi^2+a^{ij}\hat{\mathscr{D}}_l\hat{\mathscr{D}}^l\Phi^2-\hat{\mathscr{D}}^{(i}\hat{\mathscr{D}}^{j)}\Phi^2\right)
\\
\Pi^{i}_\text{m}=-\frac{1}{\Omega}\hat{\mathscr{D}}_t\Phi\hat{\mathscr{D}}^i\Phi+\frac{d-1}{4d}\left(\hat{\mathscr{D}}^i\frac{1}{\Omega}\hat{\mathscr{D}}_t\Phi^2-\hat{\mathscr{D}}_j\left(\xi^{ij}\Phi^2\right)\right)
\\
\Pi_\text{m}=\frac12\hat{\mathscr{D}}_i\Phi\hat{\mathscr{D}}^i\Phi+\frac{d-1}{4d}\left(\frac{\hat{\mathscr{R}}}{2}\Phi^2-\hat{\mathscr{D}}_{i}\hat{\mathscr{D}}^{i}\Phi^2\right).
\end{cases}
\end{eqnarray}
For the non-dynamical action, which will turn useful in a short while, we find:
\begin{equation}
\label{ndenmom}
\begin{cases}
\Pi^{ij}_\text{nd}=\frac{d-1}{4d}\left(2\varpi^{li}\varpi_{l}^{\hphantom{l}j}-\frac{a^{ij}}{2}\varpi_{lk}\varpi^{lk}\right)\Phi^2
\\
\Pi^{i}_\text{nd}=\frac{d-1}{4d}\hat{\mathscr{D}}_j\left(\varpi^{ji}\Phi^2\right)
\\
\Pi_\text{nd}=\frac{3(d-1)}{8d}\varpi_{ij}\varpi^{ij}\Phi^2.
\end{cases}
\end{equation}
They all obey \eqref{car-conf-cond}, and conservation equations \eqref{carEcon} and \eqref{carGcon} are satisfied with the electric momenta, assuming the field be on-shell, i.e. obeying \eqref{elec-eom}, and deliver the electric momentum:
\begin{equation}
 \label{elmom} 
P^i_\text{e}=\Pi^{i}_\text{m}.
\end{equation}
In a similar fashion for the magnetic dynamics, and using the equation of motion  \eqref{magn-eom}, we obtain
\begin{equation}
 \label{magmom} 
P^i_\text{m}=\Pi^{i}_\text{nd}.
\end{equation}

\lettrine[lines=2, lhang=0.33, loversize=0.25]{O}{ne might be puzzled} at this stage, by the interplay Eqs. \eqref{elmom} and \eqref{magmom} seem to entail amongst 
electric, magnetic and non-dynamics. There is no doubt that electric and magnetic Carrollian scalar dynamics resulting from $\mathcal{L}_{\text{e}}$ and $\mathcal{L}_{\text{m}}$ are distinct, and can be studied separately, on any Carrollian background. Likewise, the action $\mathcal{L}_{\text{nd}}=-V_{\text{nd}}$ is also Carrollian-invariant with bona fide Carrollian momenta, but is non-dynamical. What sets a deeper link between these dynamics, which is not visible when treating them directly in the Carrollian framework, is that they all emerge in the ``small-$c$ expansion'' of a unique relativistic theory for the scalar field. This was one possible guideline for obtaining the Carrollian scalar theories. It can also be applied to the relativistic energy--momentum tensor, and will deliver in a similar expansion\footnote{The wording ``expansion'' is an abuse because the result is exact here.} the Carrollian momenta: 
\begin{equation}
\label{em_exp}
\begin{cases}
T^{ij}=\frac{1}{c^2}\Pi_{\text{e}}^{ij}+
\Pi_{\text{m}}^{ij}+c^2 \Pi_{\text{nd}}^{ij}
\\
-\frac{c}{\Omega} T^i_{\hphantom{i}0}=
\Pi_{\text{m}}^{i}+c^2\Pi^i_{\text{nd}}
\\
\frac{1}{\Omega^2}T_{00}=\frac{1}{c^2}\Pi_{\text{e}}+
\Pi_{\text{m}}+c^2\Pi_{\text{nd}}.
\end{cases}
\end{equation}

The relationship between relativistic and Carrollian dynamics can be thrust further. Following \cite{BigFluid, CMPPS1}
we can expand the relativistic conservation of energy--momentum \eqref{conconTJ} and recollect the Carrollian conservation equations for the electric, the magnetic and the non-dynamical cases. In this process Eqs. \eqref{carEcon} and \eqref{carGcon} arise for each case at a different $c$-order, and their momenta $P^{i}_{\text{e}}$ and $P^{i}_{\text{m}}$ are naturally determined 
in terms of the next-order energy fluxes. This explains  the above results \eqref{elmom} and \eqref{magmom}. 

\lettrine[lines=2, lhang=0.33, loversize=0.25]{C}{onserved charges} are fundamental ingredients for handling a dynamical system. They often appear as the consequence of symmetries. In a relativistic framework, if $\upxi$ is a Killing field  of the spacetime $\mathscr{M}$, the current defined as 
\begin{equation}
\label{relconcur}
I_\mu=\xi^\nu T_{\mu\nu}
\end{equation}
has zero divergence and ($\mathscr{S}$ is a $d$-dimensional spatial section of $\mathscr{M}$ and $\ast\text{I}$ the  $\mathscr{M}$-Hodge dual of $\text{I} = I_\mu \text{d}x^\mu$)
\begin{equation}
\label{relconch}
Q_{I}=\int_{\mathscr{S}}\ast\text{I}
\end{equation}
is conserved. For Weyl-covariant dynamics this applies with conformal Killing fields. 

In a Carrollian spacetime a current has a scalar component $\kappa$ as well as a Carrollian-vector set of components $K^i$, and
the divergence takes the form
 \begin{equation}
\mathcal{K}= \left(
\frac{1}{\Omega}\partial_t+
\theta \right)\kappa
+\left(\hat \nabla_i +\varphi_i\right)K^i.
 \label{kilcarcon}
 \end{equation}
This result can be inferred\footnote{See Ref. \cite{BigFluid}, where it is also shown how the current components are retrieved without reference to a relativistic ascendent.} as from a relativistic computation, with a current $I^\mu$ such that 
\begin{equation}
\label{jexpCextinv}
-\frac{1}{c\Omega}I_0= \kappa + \text{O}\left(c^2\right),\quad
I^k=K^k+ \text{O}\left(c^2\right),
\end{equation}
leading in a Papapetrou--Randers background \eqref{carrp} to $\nabla_\mu I^\mu = \mathcal{K}+ \text{O}\left(c^2\right)$. 
Defining a charge associated with the current $(\kappa, \pmb{K})$ as an integral at fixed $t$ over the basis $\mathscr{S}$ of the Carrollian structure
  \begin{equation}
Q_K=\int_{\mathscr{S}}\text{d}^{d}x \sqrt{a}\left(\kappa+b_iK^i \right),
  \label{carconch}
 \end{equation}
we obtain the following time evolution:
  \begin{equation}
\frac{\text{d}Q_K}{\text{d}t}=\int_{\mathscr{S}}\text{d}^{d}x \sqrt{a}\Omega \mathcal{K}
- \int_{\partial\mathscr{S}} \star\pmb{K} \Omega,
\label{carconchdt}
 \end{equation}
 where $\star\pmb{K}$ is the $\mathscr{S}$-Hodge dual of $K_i\text{d}x^i$.  
 For vanishing divergence $\mathcal{K}$, this is conserved if one can ignore the boundary term owing to adequate fall-off or boundary conditions on the fields. Notice that if $\mathcal{K}$ happens to be identical to the Carrollian divergence of some potential $(\phi, \pmb{\phi})$, then a conserved charge is obtained with $\kappa-\phi$, $K^i-\phi^i$.

Suppose that $\upxi$ is the generator of a Carrollian diffeomorphism (see \eqref{carkil} in the appendix). It can be used to create a current out of $\Pi^{ij}$, $\Pi^{i}$,  $\Pi$ and $P^i$ \cite{CM1, BigFluid}:
\begin{equation}
\label{invkappaKcar}
\kappa= \xi^{i} P_i-\xi^{\hat t}  \Pi
,\quad K^i=\xi^{j}\Pi_{j}^{\hphantom{j}i}-
 \xi^{\hat t} \Pi^i.
\end{equation}
For a Weyl-covariant system (Eq. \eqref{car-conf-cond}) with a conformal Killing vector (see the defining conditions in the appendix, \eqref{PRkillconf} and \eqref{extra-conf-cond}), one obtains:
\begin{equation}
\label{noncons}
\mathcal{K}=
-\Pi^i\left(\left(
\hat\partial_i-\varphi_i\right)\xi^{\hat t }-
2\xi^j \varpi_{ji}
\right).
 \end{equation}
As opposed to the relativistic situation, a  conformal Killing field does not  provide a conservation law in Weyl-invariant Carrollian dynamics, unless it satisfies (the conformal weight of $\xi^{\hat t} $ is $-1$, that of $\xi^i$ zero)
\begin{equation}
\label{extraK}
\left(
\hat\partial_i-\varphi_i\right)\xi^{\hat t }-
2\xi^j \varpi_{ji}\equiv 
\hat{\mathscr{D}}_i\xi^{\hat t }-
2\xi^j \varpi_{ji}
=0.
 \end{equation}
 This last condition amounts to further demanding the clock form \eqref{kertdual} be invariant under the action of the conformal Killing (see Eq. \eqref{Liedcehrecar} in the appendix). In Carrollian dynamics, symmetry is generated by a subalgebra of the conformal isometry algebra. 

\lettrine[lines=2, lhang=0.33, loversize=0.25]{E}{lectric and magnetic Carrollian scalar fields} with conformal coupling have different behaviour regarding conservation. The former have
vanishing energy flux (see \eqref{elenmom}) and lead thus to conserved charges $Q_\text{e}=\int_{\mathscr{S}}\text{d}^{d}x \sqrt{a}\left(\kappa_\text{e}+b_iK^i_\text{e} \right)$
with
\begin{equation}
\label{invkappaKcarel}
\kappa_\text{e}= \xi^{i} \Pi_{\text{m}i}-\xi^{\hat t}  \Pi_\text{e}
,\quad K_\text{e}^i=\xi^{j}\Pi_{\text{e}j}^{i},
\end{equation}
where we have used \eqref{elmom}. For the latter, $Q_\text{m}=\int_{\mathscr{S}}\text{d}^{d}x \sqrt{a}\left(\kappa_\text{m}+b_iK^i_\text{m} \right)$ 
with (see \eqref{magenmom} and \eqref{magmom})
\begin{equation}
\label{invkappaKcarmag}
\kappa_\text{m}= \xi^{i} \Pi_{\text{nd}i}-\xi^{\hat t}  \Pi_\text{m}
,\quad K_\text{m}^i=\xi^{j}\Pi_{\text{m}j}^{i}-
 \xi^{\hat t} \Pi^i_\text{m} 
\end{equation}
is not conserved since, according to \eqref{noncons} 
\begin{equation}
 \label{noncons-mag} 
\mathcal{K}_{\text{m}}=
-\Pi^i_{\text{m}}\left(\left(
\hat\partial_i-\varphi_i\right)\xi^{\hat t }-
2\xi^j \varpi_{ji}
\right).
\end{equation}
Conservation is attainable for field configurations such that $\Pi^{i}_{m}=0$, which translates local Carroll-boost invariance \cite{Hartong:2015xda}. Following \eqref{magenmom}, this happens e.g. when $\mathscr{\hat{D}}_{t}\Phi = 0$ in backgrounds with vanishing geometric Carrollian shear ($\xi_{ij}=0$, defined in appendix, Eq. \eqref{dgammaCartime}), which is possibly compatible with the magnetic dynamics \eqref{magn-eom}.

Conformal Killing fields on general Carrollian spacetimes are obtained upon solving a set of complicated partial differential equations and this is not an easy task. It is remarkable that when the Carrollian shear (see \eqref{dgammaCartime} in the appendix) $\xi^{ij}$ vanishes, the conformal Killing fields are known \cite{Ciambelli:2019lap}. Zero shear implies that the time dependence in the metric is factorized: $a_{ij}(t,\mathbf{x})= \mathcal{B}^{-2}(t,\mathbf{x}) \tilde a_{ij}(\mathbf{x})$. This drives the conformal algebra of the Carrollian structure  to the standard infinite-dimensional semi-direct sum of the conformal algebra of $\tilde a_{ij}(\mathbf{x})$ with supertranslations. For conformally flat $\tilde a_{ij}(\mathbf{x})$, the latter coincides with
 $\mathfrak{ccarr}(d+1)$.  One recovers in particular
 $\text{BMS}_{d+2}$ in $d=1$ and $2$ --  possibly in higher dimension.\footnote{The standard conformal Carrollian algebra 
$\mathfrak{ccarr}(d+1)$ is also referred to as ``level-$2$'' $\mathfrak{ccarr}_2(d+1)$. More general level-$N$ algebras $\mathfrak{ccarr}_N(d+1)$ emerge in the presence of a dynamical exponent $z= \nicefrac{2}{N}$ 
  -- see footnote \ref{dynexp}. For  $d>2$ the BMS algebra is finite-dimensional, whereas $\mathfrak{ccarr}_{N}(d+1)$ is not. Infinite-dimensional extensions of the $\text{BMS}_{d+2}$ require adjustments in the fall-off behaviours and have been considered in the literature (see e.g. \cite{Campoleoni:2020ejn} for a recent account and further reading suggestions).} 

The Carrollian spacetimes emerging as null boundaries of asymptotically locally flat solutions to Einstein equations turn out to satisfy the vanishing-shear condition.\footnote{See e.g. \cite{Ruzziconi:2019pzd}, Eq. (3.40) at vanishing $\Lambda$ (see also \cite{CMPPS2}). One should not confuse the shear of the boundary Carrollian manifold, with the Bondi shear which is another boundary Carrollian tensor, non-vanishing in general and carrying information about the bulk gravitational radiation.} That makes this class of Carrollian structures particularly appealing and the forthcoming example will illustrate their properties regarding the propagation of a conformally coupled scalar field.

  \lettrine[lines=2, lhang=0.33, loversize=0.25]{R}{obinson--Trautman}  spacetimes are four-dimensional, time-dependent Ricci-flat solutions of algebraically special Petrov type. They describe configurations emitting gravitational radiation and settling down in the far future into a Schwarzschild black hole.\footnote{The original solution is available
in \cite{RT,Stephani:624239}. Robinson--Trautman spacetimes have been discussed in the framework of AdS/CFT in Refs. \cite{deFreitas:2014lia, Bakas:2014kfa, Gath:2015nxa, Ciambelli:2017wou}, and further in flat holography in Refs. \cite{CMPPS2, Geiller:2022vto}. \label{fRT}} Their null boundary is a Carrollian manifold $\mathscr{M}= \mathbb{R} \times \mathscr{S}$, where $ \mathscr{S}$ is equipped with a conformally flat $d=2$ metric:
\begin{equation}
\label{holoantimet}
\text{d}\ell^2=\frac{2}{P^2}\text{d}\zeta\text{d}\bar\zeta.
\end{equation}Here $P=P(t, \zeta, \bar \zeta)$ obeys a fourth-order partial-differential equation known as Robinson--Trautman's equation, which also involves the Bondi mass aspect $M(t)$.\footnote{We will not specifically use the Robinson--Trautman equation (displayed in the aforementioned literature -- footnote \ref{fRT} -- e.g. Ref. \cite{Ciambelli:2017wou}, Eq. (2.35)) in our subsequent analysis, which is thus valid for arbitrary $P(t, \zeta, \bar \zeta)$.} The field of observers and the clock form are ($\Omega=1$, $b_i = 0$)
\begin{equation}
\label{foobcl}
\upupsilon=\partial_t, \quad \upmu= -\text{d}t.
\end{equation}
Hence, one can compute the basic geometric data:\footnote{Conventions: $\sqrt{a}=\nicefrac{\text{i}}{P^2}$ and $\epsilon_{\bar\zeta\zeta}=1$.}
\begin{equation}
\label{carRT-curv} 
\theta =-2 \partial_t \ln P, \quad\varphi_i=0,\quad\varpi_{ij}=0,\quad\xi_{ij}=0,  \quad \hat{\mathscr{R}}=4P^2 \partial_{\bar\zeta} \partial_\zeta\ln P.
\end{equation}

Although the Robinson--Trautman solutions have no isometries, they have asymptotic symmetries, and these are actually reflected in the conformal isometries of the Carrollian boundary. Following \cite{Ciambelli:2019lap}, we find that the conformal Killing fields of $\mathscr{M}$ are expressed in terms of an arbitrary real function $T( \zeta, \bar \zeta)$, which encodes the \emph{supertranslations} and the conformal Killing  vectors $\pmb{Y}=Y^\zeta \partial_\zeta+Y^{\bar\zeta} \partial_{\bar\zeta}$ of $\text{d}\tilde \ell^2=2\text{d}\zeta\text{d}\bar\zeta$, which is flat space. The latter  generate $\mathfrak{so}(3,1)$ -- or even a double copy of Witt algebras referred to as \emph{superrotations}, if we are ready to give up invertibility.  We find that $\pmb{Y}$ is any combination of  $\ell_m+\bar \ell_m$ or $\text{i}\left(\ell_m-\bar \ell_m\right)$ with 
\begin{equation}
\label{lmlbarm} 
\ell_m=-\zeta^{m+1}\partial_\zeta, \quad  \bar \ell_m = -\bar\zeta^{m+1}\partial_{\bar\zeta},
\end{equation}
obeying $\text{Witt}\oplus \text{Witt}$:
\begin{equation}
\label{wittwitt} 
\left[\ell_m, \ell_n
\right]=(m-n)\ell_{m+n}, \quad
\left[\bar\ell_m,\bar \ell_n
\right]=(m-n)\bar\ell_{m+n}.
\end{equation}
In this representation, $\mathfrak{so}(3,1)$ is generated by $n=0, \pm 1$. The conformal Killing fields of $\mathscr{M}$ are (see \eqref{carkil} in the appendix)\footnote{The existence of conformal Killing fields for the Carrollian structure at hand is remarkable. Actually, the relativistic ascendent of this structure $\text{d}s^2=-c^2 \text{d}t^2+ \frac{2}{P^2}\text{d}\zeta\text{d}\bar\zeta$, appearing as the conformal time-like boundary of AdS-Robinson--Trautman spacetimes, has  generically no conformal Killings. In particular, it is not conformally flat because it has a non-zero Cotton tensor -- see\cite{deFreitas:2014lia, Bakas:2014kfa, Gath:2015nxa, Ciambelli:2017wou}.}
\begin{equation}
\label{xiYT} 
\upxi_{T,Y}
=\left(T-M_Y(C)\right) \frac{1}{P}\partial_t + Y^i\partial_i,
\end{equation}
where 
\begin{equation}
\label{C} 
C(t, \zeta, \bar \zeta)=\int^t\text{d}\tau
P(\tau, \zeta, \bar \zeta),
\end{equation}
and $M_Y$ is an operator acting on scalar functions $f(t, \zeta, \bar \zeta)$ as:
\begin{equation}
\label{MY} 
M_Y(f)=Y^k \partial_k f-\frac{f}{2}\partial_k Y^k. 
\end{equation}
The structure $\mathfrak{so}(3,1)\loplus \text{supertranslations}$ -- or $\left(\text{Witt}\oplus \text{Witt}\right)\loplus \text{supertranslations}$ -- is recovered in
\begin{equation}
\left[ \upxi_{{T}, Y},\upxi_{{T}', Y'}\right] =\upxi_{{M}_{Y}({T}')-{M}_{ Y'}({T}),[ Y, Y']}.
\end{equation}

We are now ready to  discuss the dynamics of a conformally coupled scalar field and its conserved charges. 
The ultimate motivation for this study is flat holography and the possible usefulness of the Carrollian dynamics for describing modes that propagate all the way inside the bulk towards the null boundary of asymptotically flat spacetimes. The electric equation of motion \eqref{elec-eom} reads as follows in the three-dimensional Carrollian spacetime under consideration: 
 \begin{equation}
\label{elec-eom-RT} 
\partial_t\frac{1}{P}\partial_t\frac{\Phi}{\sqrt{P}}=0.
 \end{equation}
Its general solution is given in terms of two arbitrary functions $f(\zeta, \bar \zeta)$ and 
 $g(\zeta, \bar \zeta)$:
  \begin{equation}
\label{elec-eom-RT-gensol} 
\Phi= \sqrt{P}\left(Cf+g\right).
 \end{equation}
With this, we can compute the energy density $\Pi_\text{e}$ and the electric momentum $P^i_\text{e}$ as in \eqref{elmom}, using \eqref{elenmom} and \eqref{magenmom}, and combine them into the scalar component of the current \eqref{invkappaKcarel} associated with the conformal Killing fields \eqref{xiYT}:
\begin{equation}
\label{invkappaKcarelYT}
\kappa_{\text{e}\, T,Y} 
= P^{2}\left[ Y^{i}\left(\frac{1}{4}\partial_{i}(fg) - f\partial_{i}g \right) - \frac{T f^{2}}{2} - \frac{1}{4}\partial_{i}\left( Y^{i}Cf^{2} \right) \right].
\end{equation}
This leads to the charges
\begin{equation}
\label{eleccharge}
Q_{\text{e} \, T,Y} =-\text{i} \int_{\mathscr{S}}\text{d}\zeta \wedge \text{d}\bar\zeta \left( Y^{i}\left(\frac{1}{4}\partial_{i}(fg) - f\partial_{i}g \right) - \frac{T f^{2}}{2} \right) -  \frac{1}{4}\int_{\partial\mathscr{S}} \star\pmb{Y} C f^2 P^2.
\end{equation}
On-shell, the time dependence is exclusively encoded in the last term through $P$ (and $C$). This is a flux at infinity, and thus vanishes upon appropriate fall-off behaviour of the field $f$. Hence, the charges are indeed conserved.

The infinite number of conserved charges, awkward at first glance, translates the separation of time and space imposed by Carrollian symmetry. The field equation \eqref{elec-eom-RT} contains no spatial derivative, hence every locus $(\zeta,\bar \zeta)$ provides a decoupled degree of freedom. This often happens in Carrollian field theory (as e.g. in the magnetic conformally stationary scalar field -- see below), although the general equations at hand \eqref{elec-eom} and \eqref{magn-eom} contain actually both time and space derivatives -- $\hat{\mathscr{D}}_t $ and $\hat{\mathscr{D}}_i $ contain both -- making the advertised decoupling less transparent.

The magnetic equation  \eqref{magn-eom} is 
 \begin{equation}
\label{magn-eom-RT} 
4\partial_\zeta\partial_{\bar\zeta}\Phi=\Phi \partial_\zeta\partial_{\bar\zeta}\ln P.
 \end{equation}
 According to \eqref{noncons-mag}, magnetic charges are conserved with those conformal Killing fields obeying the extra condition \eqref{extraK}, which leads to
\begin{equation}
\label{consmagkil} 
T=SP+M_Y(C), 
\end{equation}
where $S$ is a function of time only. Since $P$ and $C$ are time-dependent while $T$ isn't, Eq. \eqref{consmagkil} restricts severely the allowed subset of $\mathscr{S}$-conformal Killings $\pmb{Y}$, which may even turn empty. Assuming this set is not empty, 
due to the vanishing of the magnetic momentum $P^i_\text{m}$ (see \eqref{magmom} with \eqref{ndenmom} -- here $\varpi_{ij}=0$), Eq. \eqref{invkappaKcarmag} leads to a single conserved charge based on $\kappa_{\text{m}\,S}=-S\Pi_\text{m}$ with $\Pi_\text{m}$ given in 
\eqref{magenmom}: $Q_{\text{m}\, S}=-S\int_{\mathscr{S}}\frac{\text{d}\zeta\text{d}\bar\zeta}{P^2} \Pi_\text{m}$.
This charge is nothing but the total energy, but \emph{it turns out to vanish} here.  Indeed, on-shell, $\Pi_\text{m}$
reads (Eqs. \eqref{magn-eom} and \eqref{magenmom}), irrespective of the dimension and of the geometric  background:
\begin{equation}
\label{magenmom-en-on-shell} 
\Pi_\text{m}=\frac{1}{2d} \hat{\mathscr{D}}_{i}\left(\Phi\hat{\mathscr{D}}^{i}\Phi\right).
\end{equation}
In the case under consideration ($b_i=0$ and $\varphi_i=0$), 
$\Pi_\text{m}=\frac{P^2}{4}\left[\partial_\zeta\left(\Phi\partial_{\bar\zeta}\Phi\right)+\partial_{\bar\zeta}\left(\Phi\partial_\zeta\Phi\right)\right]$, which is a divergence. Hence $Q_{\text{m}\, S}$ receives only an $\mathscr{S}$-boundary contribution, vanishing under appropriate fall-off or boundary conditions.\footnote{This property of vanishing scalar-field conserved magnetic charges is actually valid more generally, in any dimension $d$, and for a Carrollian background structure with $b_i=0$. Indeed, this implies  $\varpi_{ij}=0$, leading therefore to 
$Q_\text{m}=-\int_{\mathscr{S}}\text{d}^{d}x \sqrt{a}\xi^{\hat t}  \Pi_\text{m}$. For Killing fields obeying the extra condition \eqref{extraK}, using \eqref{magenmom-en-on-shell} we find that the on-shell integral is again a boundary term.}

It is worth stressing that Eq. \eqref{consmagkil} is extremely constraining. For instance, if the function $P(t, \zeta, \bar \zeta)$ obeys the Robinson--Trautman equation, it can awkwardly entangle time and space dependence (see e.g. \cite{Stephani:624239}), leaving little room for finding $T(\zeta, \bar \zeta) $ and $Y^i(\zeta, \bar \zeta) $ that satisfy \eqref{consmagkil}. In the simplest possible instance, which is flat space ($P=1$ and $C=t$),\footnote{Notice in passing that the general solution of \eqref{magn-eom-RT} is in this case $\Phi(t,\zeta, \bar \zeta) = f(t, \zeta) +\bar f(t, \bar \zeta) $, where  $f(t, \zeta)$ is arbitrary.} the two special conformal transformations of the $\mathfrak{so}(3,1)$ are excluded ($\partial_i Y^i = C_0$ constant), and only constant time translations are allowed  ($T=T_0$ constant, and $S(t)=T_0 +\nicefrac{C_0 t}{2}$); this is a five-dimensional subgroup of the infinite-dimensional $\text{BMS}_{4}$.   

When $\Pi^{i}_{m}$ given in \eqref{magenmom} vanishes, the magnetic charges are all conserved, as inferred by Eq. \eqref{noncons-mag}. This occurs in particular (the Carrollian geometric shear vanishes here, see \eqref{carRT-curv}) 
 for conformally stationary scalars obeying $\frac{1}{\Omega}\hat{\mathscr{D}}_t\Phi\equiv\sqrt{P}\partial_t\frac{\Phi}{\sqrt{P}}=0$, thus of the form $\Phi=\sqrt{P} g(\zeta, \bar\zeta)$, where $g(\zeta, \bar\zeta)$ is further determined by solving the magnetic equation of motion \eqref{magn-eom-RT}. The latter\footnote{With $\Phi=\sqrt{P} g(\zeta, \bar\zeta)$, Eq.  \eqref{magn-eom-RT} reads:
 $4P\partial_\zeta \partial_{\bar \zeta} g+ 2\left(\partial_\zeta P \partial_{\bar \zeta} g +
 \partial_{\bar\zeta} P \partial_\zeta g
 \right)
 +g\partial_\zeta \partial_{\bar \zeta} P
 =0 $ (also valid if $P$ is traded for $C$).} may not be solvable in a general Robinson--Trautman background $P(t,\zeta, \bar\zeta)$ under the present ansatz. If it is,  the conserved magnetic charges are found using Eqs. \eqref{invkappaKcarmag} and  \eqref{xiYT}. On-shell, these lead to 
\begin{equation}
\label{invkappaKcarmagYT}
\kappa_{\text{m}\, T,Y} 
=-\xi^t \Pi_{\text{m}}=\frac{P^2}{2}\left(M_Y(C)-T\right)
\left(
\partial_\zeta g \partial_{\bar \zeta} g-g  \partial_\zeta \partial_{\bar \zeta} g 
\right)
,
\end{equation}
which are integrated as in
  \eqref{carconch}:
\begin{equation}
\label{magnccharge}
Q_{\text{m} \, T} =  \frac{\text{i}}{2}\int_{\mathscr{S}}\text{d}\zeta \wedge\text{d}\bar\zeta T
\left(
\partial_\zeta g \partial_{\bar \zeta} g-g  \partial_\zeta \partial_{\bar \zeta} g 
\right) -  \frac{1}{4}\int_{\partial\mathscr{S}} \star\pmb{X} P^2
\end{equation}
with
\begin{equation}
\begin{cases}
X^\zeta= C\left(Y^\zeta\left(\partial_\zeta g \partial_{\bar \zeta} g-g  \partial_\zeta \partial_{\bar \zeta} g\right)
+Y^{\bar \zeta} \left(
3\left(\partial_{\bar \zeta} g\right)^2-g \partial_{\bar \zeta}^2 g
\right)
\right)
-\frac12 Y^{\bar \zeta}  g^2 \partial_{\bar \zeta}^2C 
\\
X^{\bar\zeta}= C\left(Y^{\bar \zeta}\left(\partial_\zeta g \partial_{\bar \zeta} g-g  \partial_\zeta \partial_{\bar \zeta} g\right)
+Y^\zeta \left(
3\left(\partial_\zeta g\right)^2-g \partial_\zeta^2 g
\right)
\right)
-\frac12 Y^\zeta  g^2 \partial_\zeta^2C 

.
\end{cases}
\end{equation}
As in the electric case (see Eq. \eqref{eleccharge}), the time dependence is confined into a boundary term, which ultimately drops, taking with it all the dependence on the $\mathfrak{so}(3,1)$ vectors $\pmb{Y}$. For a conformally stationary scalar field in Robinson--Trautman background, the magnetic charges are non-zero and conserved on-shell without restriction on the Carrollian conformal Killing vector $\upxi$ (the energy flux vanishes), but they only depend on its supertranslation component $T(\zeta, \bar\zeta)$.
 
 \lettrine[lines=2, lhang=0.33, loversize=0.25]{C}{oncluding}, we would like to summarize our results. The present framework is set by a general Carrollian spacetime and the systems under investigation are general-covariant with respect to Carrollian diffeomorphisms. The Carrollian scalar field dynamics is either electric or magnetic. The same holds for a conformally coupled scalar, and the two options are rather different. The electric is ``time-like'', whereas the magnetic looks ``space-like'', and they couple to distinct pieces of the Carrollian curvature. We have determined the energy--stress tensor, the energy flux, the energy density and the momentum in both situations, and shown that Carrollian conformal isometries imply conservation laws in the electric instance but not in the magnetic.  The physical reason behind this cleavage is rather easy to understand. A Carrollian (conformal) isometry translates the invariance of the metric and the field of observers, but not that of its dual clock form. Time (supported by the field of observers) and space (associated with the clock form) directions behave differently and this ultimately reveals in the conservation properties of electric versus magnetic dynamics.  A similar phenomenon is expected to occur in Newton--Cartan manifolds, where a scalar field will also have electric and magnetic dynamics.\footnote{The magnetic and electric  scalar-field actions are respectively $S_{\text{m}} = -\int_\mathscr{M}  \text{d}t \,\text{d}^{d}x \sqrt{a} \Omega 
 \left( \frac{1}{2} a^{ij}\partial_{i}\Phi \partial_{j}\Phi + V_{\text{m}} (\Phi) \right)
$
and
$ S_{\text{e}} = \int_\mathscr{M}  \text{d}t \,\text{d}^{d}x \sqrt{a} \Omega 
 \left( \left(\frac{1}{\Omega}\frac{\hat{\text{D}}\Phi}{\text{d}t} \right)^{2} - V_{\text{e}} (\Phi) \right)$ in torsionless Newton--Cartan geometries with degenerate cometric $a^{ij}$, clock form $\Omega \text{d}t$, field of observers $\frac{1}{\Omega} \left(\partial_t +w^j\partial_j\right)$, and metric-compatible time derivative
$\frac{1}{\Omega}\frac{\hat{\text{D}}\Phi}{\text{d}t}
= \frac{1}{\Omega} \partial_t\Phi+\frac{w^{ j}}{\Omega}\partial_j\Phi 
$.  }  Isometries will guarantee conservation laws for the latter, as opposed to the former, because the clock form is invariant under the action of a Killing vector, while the field of observers isn't.
 
The above findings have been illustrated in the case of the null boundary of Robinson--Trautman asymptotically locally flat spacetimes, which are Carrollian with vanishing geometric shear and vorticity. The electric conformally coupled scalar field has been worked out thoroughly, accompanied with its infinite tower of conserved charges. For the magnetic dynamics, we have found that all charges associated with the subalgebra of the conformal Carrollian algebra satisfying the extra conservation condition (Eq.~\eqref{extraK}) vanish -- i.e. amount to purely boundary terms. Non-vanishing conserved magnetic charges appear for field configurations with  $\Pi^{i}_{m}=0$, and this happens e.g. for conformally stationary fields. 

From our general discussion one should probably retain the contrast between the infinite tower of conformal Killing fields available in most Carrollian structures and the often lesser conserved Carrollian charges. In this picture one should not underestimate the role of the non-conserved ones, usually infinite in number. When the Carrollian structures are null boundaries of asymptotically flat spacetimes, the presence of non-conserved charges betrays, among others, gravitational radiation. 
 
Even though we have focused our analysis on conformally coupled scalar fields, ordinary scalars share these properties -- with Killings instead of conformal Killings. The motivation behind conformal couplings  lies in the role these may play in flat holography -- for scalar or more general fields. This calls for a better understanding of the classical dynamics, and above all of the quantum properties. The conservation of  charges, the associated algebras and the distinction of electric versus magnetic representatives, should ultimately be translated into bulk language. Our  example of the null three-dimensional boundary of Robinson--Trautman Ricci-flat spacetimes in meant to illustrate this bridge, although discussed here in a primitive fashion, revealing a generically  trivial magnetic conservation as opposed to an infinite set of electric  conserved charges. How this reflects flat-holographic properties remains in limbo. 
  
 \section*{Acknowledgements}

We would like to express our gratitude to our colleagues Marios Petropoulos and Konstantinos Siampos for carefully reading the draft of this manuscript, for many useful discussions and for suggestions of improvements. We also thank Jelle Hartong and Gerben Oling for interesting exchanges on Carrollian covariance and Carroll-boost invariance.
The work of D. Rivera-Betancour was funded by Becas Chile (ANID) Scholarship No. 72200301. 
The work of M. Vilatte  was supported by the Hellenic Foundation for Research and Innovation (H.F.R.I.) under the \textsl{First Call for H.F.R.I. Research Projects to support Faculty members and Researchers and the procurement of high-cost research equipment grant} (MIS 1524, Project Number: 96048).

\appendix 

\section*{Appendix: Carrollian manifolds}

\lettrine[lines=2, lhang=0.33, loversize=0.25]{U}{nder Carrollian diffeomorphisms} \eqref{cardifs} with Jacobian 
\begin{equation}
 \label{carj}
J(t,\mathbf{x})=\frac{\partial t'}{\partial t},\quad j_i(t,\mathbf{x}) = \frac{\partial  t'}{\partial x^{i}},\quad 
J^i_j(\mathbf{x}) = \frac{\partial x^{i\prime}}{\partial x^{j}},
\end{equation}
the transformations are non-covariant (connection-like) for $\partial_i$ and $b_i$, and density-like for $\partial_t$ and $\Omega$: 
\begin{equation}
\partial^\prime_j=J^{-1i}_{\hphantom{-1}j}\left(\partial_i-
\frac{j_i}{J}\partial_t\right),\quad b^{\prime}_{k}=\left( b_i+\frac{\Omega}{J} j_i\right)J^{-1i}_{\hphantom{-1}k},\quad 
\partial^\prime_t=\frac{1}{J}\partial_t, \quad
\Omega^{\prime }=\frac{\Omega}{J}.
\end{equation}
They are covariant for the other objects:
\begin{equation}
\upupsilon^{\prime}
=
\upupsilon,\quad
\upmu^{\prime}
=
\upmu,\quad
\hat\partial_i^\prime =
J^{-1j}_{\hphantom{-1}i} \hat\partial_j, \quad
a^{ ij\prime}=J^i_k J_l^ja^{kl}.
\end{equation}

Carrollian tensors depend on time $t$ and space $\mathbf{x}$, carry indices $i, j, \ldots $  lowered and raised with $a_{ij}$ and its inverse $a^{ij}$, and transform covariantly under Carrollian diffeomorphisms \eqref{cardifs} with  $J_i^j$ and $J^{-1i}_{\hphantom{-1}j}$ defined in \eqref{carj}. A Levi--Civita--Carroll spatial covariant derivative $\hat \nabla_i$
is defined with connection coefficients 
\begin{equation}
\label{dgammaCar}
\hat\gamma^i_{jk}=\dfrac{a^{il}}{2}\left(
\hat\partial_j
a_{lk}+\hat\partial_k  a_{lj}-
\hat\partial_l a_{jk}\right),
\end{equation}
which emerge naturally in the vanishing-$c$ limit of a Levi--Civita connection  
in the Papapetrou--Randers coordinates \eqref{carrp}. This connection is torsionless and metric-compatible:\footnote{Details on the transformation rules can be found in the appendix A.2 of Ref. \cite{CMPPS1}, together with further useful properties.}
$\hat\gamma^k_{[ij]}=0$,  $\hat\nabla_ia_{jk}=0$.
The vectors $\hat\partial_i$ do not commute and define the Carrollian vorticity and acceleration:
\begin{equation}
\label{carconcomderf}
\left[\hat\partial_i,\hat\partial_j\right]=
\frac{2}{\Omega}\varpi_{ij}\partial_t,\quad\varpi_{ij}=\partial_{[i}b_{j]}+b_{[i}\varphi_{j]},\quad 
\varphi_i=\dfrac{1}{\Omega}\left(\partial_t b_i+\partial_i \Omega\right).
\end{equation}

The usual time-derivative operator $\frac{1}{\Omega}\partial_t$ acts covariantly on Carrollian tensors, but it is not metric-compatible because $a_{ij}$ depend on time. A  temporal covariant derivative is defined by requiring  $\frac{1}{\Omega^\prime}{\hat D}^\prime_t=\frac{1}{\Omega}\hat D_t$ and $\hat D_ta_{jk}=0$,
and is also inherited from the Papapetrou--Randers  Levi--Civita connection. To this end,
we introduce a temporal connection
\begin{equation}
\label{dgammaCartime}
\hat\gamma_{ij}=\frac{1}{2\Omega}\partial_t a_{ij}
=\xi_{ij} + \frac{1}{d}a_{ij}\theta,\quad \theta=
\dfrac{1}{\Omega}              
\partial_t \ln\sqrt{a},
\end{equation}
which is a  symmetric Carrollian tensor split into the Carrollian shear (traceless) and the Carrollian expansion (trace). 
The action of  $\hat D_t$ on scalars is $\partial_t$
whereas on vectors or forms it
is defined as
\begin{equation}
\label{Cartimecovdervecform}
\frac{1}{\Omega}\hat D_tV^i=\frac{1}{\Omega} \partial_tV^i+\hat\gamma^i_{\hphantom{i}j} V^j,\quad
\frac{1}{\Omega}\hat D_tV_i=\frac{1}{\Omega} \partial_tV_i-\hat\gamma_i^{\hphantom{i}j} V_j.
\end{equation}
Generalization to any tensor uses Leibniz rule.

The commutators of Carrollian covariant derivatives define Carrollian curvature tensors. We keep it minimal here with  
\begin{equation}
\begin{array}{rcl}
\left[\hat\nabla_k,\hat\nabla_l\right]V^i&=&\left(
\hat\partial_k\hat\gamma^i_{lj}
-\hat\partial_l\hat\gamma^i_{kj}
+\hat\gamma^i_{km}\hat\gamma^m_{lj}
-\hat\gamma^i_{lm}\hat\gamma^m_{kj}
\right)V^j+\left[\hat\partial_k,\hat\partial_l\right]V^i
\\
&=& \hat r^i_{\hphantom{i}jkl}V^j+
\varpi_{kl}\frac{2}{\Omega}\partial_{t}V^i.
\end{array}
\label{carriemann}
\end{equation}
In this expression $ \hat r^i_{\hphantom{i}jkl}$ should be called the \emph{Riemann--Carroll} tensor. The Ricci--Carroll tensor and the Carroll scalar curvature are thus
\begin{equation}
\label{carricci-scalar}
\hat r_{ij}=\hat r^k_{\hphantom{k}ikj}\neq \hat r_{ji},\quad \hat r=a^{ij}\hat r_{ij}.
\end{equation}

\lettrine[lines=2, lhang=0.33, loversize=0.25]{W}{eyl-covariance} under  Weyl transformations
\begin{equation}
\label{weyl-geometry-abs}
a_{ij}\to \frac{1}{\mathcal{B}^2}a_{ij},\quad b_{i}\to \frac{1}{\mathcal{B}}b_{i},\quad \Omega\to \frac{1}{\mathcal{B}}\Omega
\end{equation} 
with $\mathcal{B}=\mathcal{B}(t,\mathbf{x})$ an arbitrary function, requires a \emph{Weyl--Carroll} connection built on $\varphi_i$ and  $\theta$ defined in \eqref{carconcomderf}
and
\eqref{dgammaCartime}, 
which transform as 
 \begin{equation}
 \label{weyl-geometry-2-abs}
 \varphi_{i}\to \varphi_{i}-\hat\partial_i\ln \mathcal{B}
  ,\quad 
\theta\to \mathcal{B}\theta-\frac{d}{\Omega}\partial_t \mathcal{B}.
\end{equation} 
The Carrollian vorticity $\varpi_{ij}$ \eqref{carconcomderf} and the Carrollian 
shear $\xi_{ij}$ \eqref{dgammaCartime} are Weyl-covariant of weight $-1$. 

The Weyl--Carroll space and time covariant derivatives are torsionless and metric-compatible.
For a weight-$w$ scalar function  $\Phi$ and  a vector with weight-$w$ components $V^l$, the action is
\begin{eqnarray}
\label{CWs-Phi}
&&\hat{\mathscr{D}}_j \Phi=\hat\partial_j \Phi +w \varphi_j \Phi,\\
&&\hat{\mathscr{D}}_j V^l=\hat\nabla_j V^l +(w-1) \varphi_j V^l +\varphi^l V_j -\delta^l_j V^i\varphi_i.
\end{eqnarray}
The Weyl--Carroll spatial derivative does not alter the weight, and one checks that $\hat{\mathscr{D}}_j a_{kl}=0$.
Regarding time, one  defines 
\begin{eqnarray}
\label{CWtimecovdersc}
&&\frac{1}{\Omega}\hat{\mathscr{D}}_t \Phi=\frac{1}{\Omega}\hat D_t \Phi +\frac{w}{d} \theta \Phi=
\frac{1}{\Omega}\partial_t \Phi +\frac{w}{d} \theta \Phi,\\
&&\frac{1}{\Omega}\hat{\mathscr{D}}_t V^l=\frac{1}{\Omega}\hat D_t V^l +\frac{w-1}{d} \theta V^l
=\frac{1}{\Omega}\partial_t V^l +\frac{w}{d} \theta V^l+\xi^{l}_{\hphantom{l}i} V^i ,
\label{CWtimecovdervecform}
\end{eqnarray}
and both are of weight $w+1$. Similarly for any tensor by Leibniz rule and in particular we find
$\hat{\mathscr{D}}_t a_{kl}=0$.

We now close this paragraph with the Weyl--Carroll curvature tensors, appearing in the commutation of Weyl--Carroll covariant derivatives. We find 
\begin{eqnarray}
\label{CWcontor}
&&\left[\hat{\mathscr{D}}_i,\hat{\mathscr{D}}_j\right]\Phi=
\frac{2}{\Omega}\varpi_{ij}\hat{\mathscr{D}}_t\Phi
+w \Omega_{ij} 
\Phi,
\\
\label{CWcurvten}
&&\left[\hat{\mathscr{D}}_k,\hat{\mathscr{D}}_l\right]V^i=
\left( \hat{\mathscr{R}}^i_{\hphantom{i}jkl} - 2
\xi^{i}_{\hphantom{i}j}
\varpi_{kl} 
\right)
V^j+
\varpi_{kl}\frac{2}{\Omega}\hat{\mathscr{D}}_t V^i
+w \Omega_{kl}
V^i,
\end{eqnarray}
where we have introduced the following Carrollian, weight-$0$ Weyl-covariant tensors: 
\begin{eqnarray}
\label{CWRiemann}
\hat{\mathscr{R}}^i_{\hphantom{i}jkl} &=&\hat r^i_{\hphantom{i}jkl}
-\delta^i_j\varphi_{kl}
-a_{jk} \hat{\nabla}_l \varphi^i
+a_{jl} \hat{\nabla}_k \varphi^i 
+\delta^i_k \hat{\nabla}_l \varphi_j 
-\delta^i_l \hat{\nabla}_k \varphi_j 
\nonumber
\\ 
&&+\varphi^i\left(\varphi_k a_{jl}-\varphi_l a_{jk}\right)
-\left(\delta^i_k a_{jl}-\delta^i_l a_{jk}\right)\varphi_m\varphi^m+
\left(\delta^i_k \varphi_l-\delta^i_l \varphi_k\right)\varphi_j,\\
\label{CWOme}
 \Omega_{ij} &=&\hat\partial_i \varphi_j - \hat\partial_j \varphi_i -\frac{2}{d}\varpi_{ij} \theta.
 \end{eqnarray}
Additionally, we define traces as:
\begin{equation}
\label{CWricci-scalar}
\hat{\mathscr{R}}_{ij}=\hat{\mathscr{R}}^k_{\hphantom{k}ikj}
,\quad \hat{\mathscr{R}}=a^{ij}\hat{\mathscr{R}}_{ij}
\end{equation}
with
\begin{equation}
\label{CWscalar}
\hat{\mathscr{R}}=\hat r +(d-1)\left(2\hat \nabla_i\varphi^i-(d-2)\varphi_i\varphi^i\right).
\end{equation}
Observe that the Weyl-covariant Carroll--Ricci tensor is not symmetric: $\hat{\mathscr{R}}_{[ij]}=-\frac{d}{2} \Omega_{ij}$.
Finally, we recall that
\begin{equation}
\left[\frac{1}{\Omega}\hat{\mathscr{D}}_{t},\hat{\mathscr{D}}_i\right]\Phi= w \hat{\mathscr{R}}_{i}\Phi-
\xi^{j}_{\hphantom{j}i}\hat{\mathscr{D}}_j^{\vphantom{j}} \Phi,
\label{CWrsc}
\end{equation}
where
\begin{equation}
\hat{\mathscr{R}}_{i}=
\frac{1}{\Omega} \partial_{t}\varphi_i-\frac{1}{d}\left(\hat \partial_i+\varphi_i\right)\theta
\label{CWRvec}
\end{equation}
are the components of a Weyl-covariant weight-$1$ Carrollian curvature one-form.

\lettrine[lines=2, lhang=0.33, loversize=0.25]{I}{sometries and conformal isometries} are associated with Killing and conformal Killing fields.
Carrollian diffeomorphisms \eqref{cardifs} are generated by vector fields
\begin{equation}   
\label{carkil}
\upxi=\xi^t \partial_t +\xi^i \partial_i= \left(\xi^t -\xi^i\frac{b_i}{\Omega}\right) \partial_t
+ \xi^i \left(\partial_i+\frac{b_i}{\Omega}\partial_t\right)=
\xi^{\hat t}\frac{1}{\Omega} \partial_t+\xi^i \hat \partial_i 
 \end{equation}
restricted to $\xi^i=\xi^i(\mathbf{x})$. Their action operates with the Lie derivative, and for the geometric data one finds
\begin{eqnarray}   
\label{Liedaijcar}
\mathscr{L}_\upxi a_{ij}&=& 2\hat\nabla_{(i}\xi^ka_{j)k}+2\xi^{\hat t}  \hat \gamma_{ij}
,\\
\label{Liedcfoocar}
\mathscr{L}_\upxi \upupsilon&=&
\mu
\upupsilon
,\\
\label{Liedcehrecar}
\mathscr{L}_\upxi \upmu&=&-\mu  \upmu -\left(\left(\hat\partial_i-\varphi_i\right)\xi^{\hat t }-
2\xi^j \varpi_{ji}\right)\text{d}x^i
\end{eqnarray}
with
\begin{equation}
\mu(t,\mathbf{x})= -\left(\frac{1}{\Omega} \partial_t
\xi^{\hat t }+
\varphi_i \xi^i \right) .
 \end{equation}
The significant observation is here that due to the degeneration of the metric on $\mathscr{M}$,  \emph{the variation of the clock form $\upmu$ is not identical to that of the field of observers $\upupsilon$}. For further use, we also introduce the trace of \eqref{Liedaijcar} devided by $d$:
\begin{equation}
\label{carkilleqconf}
\lambda(t,\mathbf{x}) =\frac{2}{d}
\left(\hat  \nabla_{i}\xi^{i}
+\theta \xi^{\hat t}  
\right).
 \end{equation}

Carrollian isometries are Carrollian diffeomorphisms generated by Killing fields, obeying $\mathscr{L}_\upxi a_{ij}=0$ and $\mathscr{L}_\upxi \upupsilon=0$.
In the strong  Carroll structure, this requirement is completed with the invariance of the connection. 
For conformal Carrollian isometries one demands 
\begin{equation}
\label{PRkillconf}
\mathscr{L}_\upxi a_{ij}=\lambda a_{ij}.
\end{equation}
This set of partial differential equations is insufficient for defining conformal Killing vectors and one usually imposes to tune $\mu$ versus $\lambda$ so that the scaling of the metric be twice that of the field of observers:\footnote{One usually considers in the literature $2\mu + z\lambda=0$,
where $z$ is minus the conformal weight of $\Omega$, referred to as the \emph{dynamical exponent}. Here, due to the relationship of the considered Carrollian spacetimes with relativistic parents,  the weight of  $\Omega$ is $-1$. \label{dynexp}}

\begin{equation}
\label{extra-conf-cond}
2\mu + \lambda=0.
 \end{equation}
The projective structure associated with some Carroll connection should also be preserved.

\lettrine[lines=2, lhang=0.33, loversize=0.25]{F}{or a pseudo-Riemannian manifold}  $\mathscr{M}$ in $d+1$ dimensions with metric $g_{\mu\nu}$ (weight-$2$),  
a Weyl-covariant derivative $\mathscr{D}_\mu$ maintains the weight $w$ of a Weyl-covariant tensor. The corresponding connection uses a  (weight-$1$)  vector $u^\mu$ of norm $-c^2$, as well as its expansion $\Theta$ and acceleration $a^\mu$:
\begin{equation}
\label{Wconc}
\text{A}=\frac{1}{c^2}\left(\text{a} -\frac{\Theta}{d} \text{u}\right).
\end{equation}
The Weyl covariant derivative is metric-compatible with 
\begin{equation}
\left(\mathscr{D}_\mu\mathscr{D}_\nu -\mathscr{D}_\nu\mathscr{D}_\mu\right) f= w f F_{\mu\nu},\quad F_{\mu\nu}=\partial_\mu A_\nu-\partial_\nu A_\mu,
\end{equation}
where  the action on a weight-$w$ scalar $f$ is
\begin{equation}
\label{WCscalar}
\mathscr{D}_\lambda f=\nabla_\lambda  f +w A_\lambda f.
\end{equation}
The action of $\mathscr{D}_\lambda$  on a weight-$w$ form $v_\mu$  is
\begin{equation}
\label{Wv}
\mathscr{D}_\lambda v_\mu=\nabla_\lambda v_\mu+(w+1)A_\lambda v_\mu + A_\mu v_\lambda-g_{\mu\lambda} A^\rho v_\rho,
\end{equation}
and we obtain 
\begin{equation}
\label{relweylcurv}
\left(\mathscr{D}_\mu\mathscr{D}_\nu -\mathscr{D}_\nu\mathscr{D}_\mu\right) v^\rho=
\mathscr{R}^\rho_{\hphantom{\rho}\sigma\mu\nu} v^\sigma+ w v^\rho F_{\mu\nu}.
\end{equation}
The Weyl-covariant Ricci (weight $0$) and scalar (weight $2$) curvatures read: 
\begin{eqnarray}
\mathscr{R}_{\mu\nu}&=&{R}_{\mu\nu} + (d-1)\left(\nabla_\nu A_\mu + A_\mu A_\nu-g_{\mu\nu}A_\lambda A^\lambda\right) +
g_{\mu\nu}\nabla_\lambda A^\lambda
-F_{\mu\nu},
\label{curlRic}
\\
\mathscr{R}&=&R +2d\nabla_\lambda A^\lambda- d(d-1) A_\lambda A^\lambda . 
\label{curlRc}
\end{eqnarray}
Observe that $\mathscr{R}_{\mu\nu}$ is not symmetric.

If the metric is of the Papapetrou--Randers form \eqref{carrp}, the dependence with respect to the velocity of light $c$ is explicit. Thus every relativistic tensor, i.e. a tensor with respect to the full diffeomorphism group, can be reduced with respect to the Carrollian subgroup \eqref{cardifs}, and exhibits a finite number of Carrollian tensors. We find:
\begin{eqnarray}
\label{reducR}
R&=&\frac{1}{c^2}\left(\frac{2}{\Omega}\partial_t \theta+\frac{1+d}{d}\theta^2+\xi_{ij}\xi^{ij}\right)+\hat{r}-2\hat{\nabla}_{i}\varphi^{i}-2\varphi^{i}\varphi_{i}+c^2\varpi_{ij}\varpi^{ij},
\\
\label{reduccurlR}
\mathscr{R}&=&\frac{1}{c^2}\xi_{ij}\xi^{ij}+\hat{\mathscr{R}}+c^2\varpi_{ij}\varpi^{ij}.
\end{eqnarray}
Actually, the Carroll and Weyl--Carroll connections introduced earlier are also obtained from the ordinary Levy--Civita and the Weyl connections of the pseudo-Riemannian spacetaime at hand, in the form of an exact, truncated Laurent expansion.

\end{document}